\documentclass[conference]{IEEEtran}
\IEEEoverridecommandlockouts
\usepackage{cite}
\usepackage{amsmath,amssymb,amsfonts}
\usepackage{algorithmic}
\usepackage{graphicx}
\usepackage{textcomp}
\usepackage{xcolor}
\usepackage{amssymb}
\usepackage{amsmath}
\usepackage{graphicx}
\usepackage{epstopdf}
\usepackage{balance}
\usepackage{epsfig}
\usepackage{graphics}
\usepackage{subfigure}
\usepackage{lineno}
\usepackage{epsfig}
\usepackage{color}
\usepackage{textcomp}
\usepackage{algorithm}  
\usepackage{algorithmic} 
\usepackage{bm}
\usepackage{upgreek}
\graphicspath{ {Figures/}}
\usepackage{setspace}
\begin{document}

\title{
\begin{spacing}{0.4}
Waveform Design for Communication-Assisted Sensing in 6G Perceptive Networks
\end{spacing}
}

\author{\IEEEauthorblockN{Fuwang Dong$^1$, Fan Liu$^{1,*}$, Shihang Lu$^1$, Weijie Yuan$^1$, Yuanhao Cui$^1$, Yifeng Xiong$^{1,2}$, Feifei Gao$^3$}
\IEEEauthorblockA{
%\textit{$^1$College of Intelligent Systems Science and Engineering, Harbin Engineering University, Harbin, China}\\
\textit{$^1$Department of Electronic and Electrical Engineering, Southern University of Science and Technology, Shenzhen, China}\\
\textit{$^2$School of Information and Communication Engineering, Beijing University of Posts and Telecommunications, Beijing, China}\\
\textit{$^3$Department of Automation, Tsinghua University, Beijing, China}\\
E-mail: \{dongfw, liuf6\}@sustech.edu.cn; lush2021@mail.sustech.edu.cn; \\
\{yuanwj, cuiyh\}@sustech.edu.cn; yifengxiong@bupt.edu.cn; feifeigao@ieee.org}
}

\maketitle

\begin{abstract}
The integrated sensing and communication (ISAC) technique has the potential to achieve coordination gain by exploiting the mutual assistance between sensing and communication (S$\&$C) functions. While the sensing-assisted communications (SAC) technology has been extensively studied for high-mobility scenarios, the communication-assisted sensing (CAS) counterpart remains widely unexplored. This paper presents a waveform design framework for CAS in 6G perceptive networks, aiming to attain an optimal sensing quality of service (QoS) at the user after the target's parameters successively ``pass-through'' the S$\&$C channels. In particular, a pair of transmission schemes, namely, separated S$\&$C and dual-functional waveform designs, are proposed to optimize the sensing QoS under the constraints of the rate-distortion and power budget. The first scheme reveals a \textit{power allocation} trade-off, while the latter presents a \textit{water-filling} trade-off. Numerical results demonstrate the effectiveness of the proposed algorithms, where the dual-functional scheme exhibits approximately 25\% performance gain compared to its separated waveform design counterpart.         
\end{abstract}


\section{Introduction}\label{Introduction}
Next-generation wireless networks (5G-A and 6G) are envisioned to simultaneously provide high-precision sensing capabilities and robust wireless connectivity through the integrated sensing and communication (ISAC) technology, leading to the emergence of perceptive mobile networks \cite{9296833}. ISAC techniques can offer two potential gains, integration gain and coordination gain \cite{9737357}, as opposed to traditional separate deployment.  The integration gain has been extensively studied in the past few decades, including the sharing of wireless resources and hardware platforms through sophisticated designs of ISAC waveforms, beamforming, and transmission protocols, etc. (cf.\cite{8999605,9540344}, and the reference therein). However, there is relatively little literature on how to exploit the coordination gain in ISAC systems.

Coordination gains are often achieved through the collaboration of sensing and communication (S$\&$C) functions. The existing sensing-assisted communication (SAC) techniques aim to reduce pilot training overhead in channel estimation by leveraging radar sensing capabilities. In \cite{7929295}, an auxiliary beam pair method is used in mmWave communication systems to estimate departure/arrival angles by comparing auxiliary beam pair amplitudes. Recently, the authors of \cite{9171304} proposed an SAC framework for vehicular networks based on extended Kalman filtering, which significantly reduces communication overheads by sophisticatedly tailored beam tracking and prediction approaches. Later, the SAC performance was further improved by adopting a message passing algorithm \cite{9246715}. While SAC techniques have been well-studied for high-mobility communication systems, it remains unclear how to improve the sensing performance by employing the communication functionality. 

In conventional wireless sensor networks, the communication function has already been employed to deliver sensing data gathered by other sensors. However, the working pipeline for CAS techniques in perceptive networks differs significantly. Let us consider the scenario where the users require to sense targets of interest beyond their line of sight (LoS), or to obtain more accurate observations of long-range targets. In the CAS framework, the nearby base station (BS) with favorable views can help to extract the targets' parameters via device-free wireless sensing without the assistance of additional sensors, and communicate the sensory information to the users simultaneously. Accordingly, the impact of S$\&$C procedures on the sensing quality of service (QoS) attained by the users and the competition for system resources at the BS lead to unique challenges and opportunities in CAS system design.  
 
In this paper, we commence with establishing a general transmit waveform design framework for CAS systems. Then, we propose two transmission schemes: the dual-functional and separated S$\&$C waveform designs, that optimize sensing QoS under the constraints of source-channel separation and total power budget, respectively. We derive the explicit expressions for performance metrics and the rate-distortion (R-D) function in the scenarios of target response matrix (TRM) estimation. The waveform design problem is then transformed to power allocation with respect to the eigenvalues of the signal covariance matrices. Finally, we present the solution algorithms and validate our findings through numerical simulations.  
 
\begin{figure*}[!t]
	\centering
	\includegraphics[width=5in]{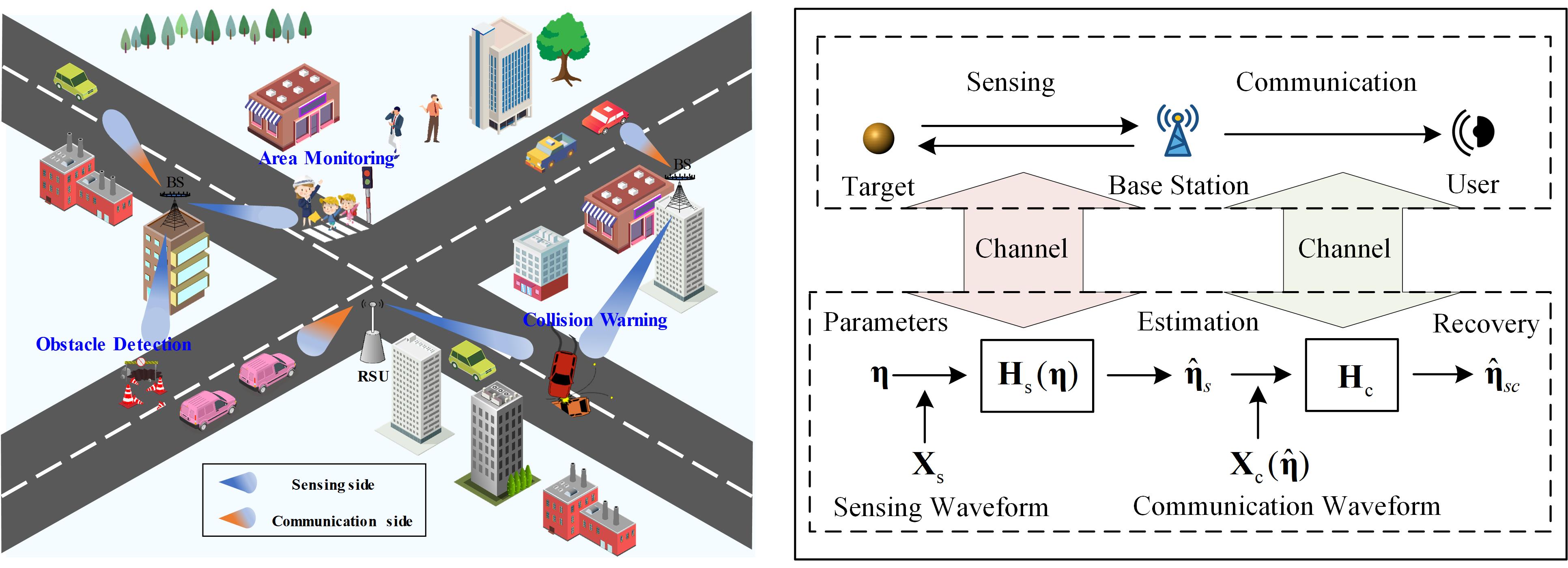}
	\caption{The typical use cases of the CAS technique and the system model.}
	\label{fig1}
\end{figure*}

\section{System Model}\label{SystemModel}
\subsection{The CAS Framework}
The typical use cases of CAS are shown in Fig. \ref{fig1}, where a user wishes to sense a target within its Non-LoS (NLoS) paths, and hence requests sensing service from the nearby BS, who observes the target of interest from a LoS path. Let us denote $\bm{\upeta} \in \mathbb{R}^K$ as the parameter vector to be sensed, such as range, azimuth, TRM, etc., which takes value on a set $\mathcal{A}$ with a prior distribution $p_{\bm{\upeta}}(\bm{\eta})$. The procedure that the BS senses the target and transmits the estimated results to the user, can be equivalently considered as that the target information successively ``passes through'' the sensing and communication channels before its arrival at the user. In such a case, the original $\bm{\upeta}$, the estimated $\hat{\bm{\upeta}}_s$ at the BS, and the recovered $\hat{\bm{\upeta}}_{sc}$ at the user forms a Markov chain $\bm{\upeta} \to \hat{\bm{\upeta}}_s \to \hat{\bm{\upeta}}_{sc}$. The detailed signal models are as follows.

$\bullet$ $\textbf{Sensing side (S-side)}$: The BS transmits sensing waveform to the target and yields the estimated parameter, namely, $\hat{\bm{\upeta}}_s$ through the noisy received echo signals with the expression of 
\begin{equation}\label{SensingM}
\textbf{Y}_s=\textbf{H}_s(\bm{\upeta})\textbf{X}_s+\textbf{Z}_s,
\end{equation} 
where $\textbf{X}_s \in \mathbb{C}^{N \times T}$ is the transmitted sensing waveform. $\textbf{H}_s(\bm{\upeta}) \in \mathbb{C}^{M_s \times N}$ represents the sensing channel matrix with respect to (w.r.t.) the latent parameters $\bm{\upeta}$. $\textbf{Z}_s \in \mathbb{C}^{M_s \times T}$ is the additive noise, whose entries follow an independently and identically circularly symmetric complex Gaussian distribution with $\mathcal{CN}(0,\sigma_s^2\textbf{I}_{TM_s})$. $T$, $N$ and $M_s$ are the numbers of symbols, transmitting and receiving antennas at the BS, respectively. 

$\bullet$ $\textbf{Communication side (C-side)}$: The BS transmits the estimated information $\hat{\bm{\upeta}}_s$ to the user through source-channel coding, and the user reconstruct an estimate $\hat{\bm{\upeta}}_{sc}$ from the following received signal
\begin{equation}
\textbf{Y}_c=\textbf{H}_c\textbf{X}_c(\hat{\bm{\upeta}}_s)+\textbf{Z}_c,
\end{equation} 
where $\textbf{X}_c(\hat{\bm{\upeta}}_s) \in \mathbb{C}^{N \times T}$ is the transmitted communication waveform carrying the information of $\hat{\bm{\upeta}}_s$. $\textbf{H}_c \in \mathbb{C}^{M_c \times N}$ represents the communication channel matrix. $\textbf{Z}_c \in \mathbb{C}^{M_c \times T}$ follows similar distribution as to $\textbf{Z}_s$ with variance $\sigma_c^2$. $M_c$ is the number of the equipped antennas at the user.  

A reasonable metric for sensing QoS is the mean square error (MSE) between the recovery $\hat{\bm{\upeta}}_{sc}$ at the user and the ground truth $\bm{\upeta}$. Then the average distortion can be defined by  
\begin{equation} {\label{Dsc}} 
D_{sc}=\mathbb{E}_{\bm{\upeta}}[\left\|\bm{\upeta}-\hat{\bm{\upeta}}_{sc}\right\|_2^2].
\end{equation}
We adopt the minimum MSE (MMSE) estimator at S-side, i.e., $\hat{\bm{\upeta}}_{s} = \mathbb{E}_{\bm{\upeta}}\left[ \bm{\upeta}|\textbf{Y}_s\right]$, the distortion $D_{sc}$ can be recast by  
\begin{equation} {\label{Dsc2}} 
\begin{aligned}	
D_{sc}&=\mathbb{E}_{\bm{\upeta}}\left[\left\|\bm{\upeta}-\hat{\bm{\upeta}}_{s}+\hat{\bm{\upeta}}_{s}-\hat{\bm{\upeta}}_{sc}\right\|_2^2\right] \\
&\mathop = \limits^{(a)} \mathbb{E}_{\bm{\upeta}}\left[\left\|\bm{\upeta}-\hat{\bm{\upeta}}_{s}\right\|_2^2 \right]+\mathbb{E}_{\bm{\upeta}}\left[\left\|\hat{\bm{\upeta}}_s-\hat{\bm{\upeta}}_{sc}\right\|_2^2 \right] \mathop = \limits^{\Delta} D_s+D_c,
\end{aligned}
\end{equation}
where $(a)$ holds from the properties of the conditional expectation $\mathbb{E}\left[(\bm{\upeta}-\mathbb{E}\left[ \bm{\upeta}|\textbf{Y}_s\right])^Tf(\textbf{Y}_s)\right]=0$. In (\ref{Dsc2}), we demonstrate that the sensing QoS depends on both the estimation distortion $D_s$ at S-side and the recovery distortion $D_c$ at C-side. Moreover, it is worth highlighting that the above CAS process is to some extent similar to the \textit{remote source coding} (also known as the \textit{CEO problem}) in information theory, where the encoder can only access a noisy observation rather than the original source of interest \cite{8636539}.

\subsection{Problem Formulation}
Evidently, the distortion $D_s$ and $D_c$ depends on the transmitting waveform $\textbf{X}_s$ and $\textbf{X}_c$, respectively, which motivates us to design the optimal waveform such that the distortion $D_{sc}$ achieved at the user is minimized. Before presenting the formulation, we briefly introduce the rate-distortion (R-D) theory to characterize the relationship between the distortion $D_c$ and the channel capacity at C-side. For a random variable $\bm{\upeta}$ and its estimate $\hat{\bm{\upeta}}$, the R-D function is given by \cite{thomas2006elements}   
\begin{equation}\label{ER}
	R(D)=\mathop { \rm{min} } \limits_{P_{\hat{\bm{\upeta}}|\bm{\upeta}}:\mathbb{E}\left[d(\bm{\upeta},\hat{\bm{\upeta}})\right] \le D} I(\bm{\upeta};\hat{\bm{\upeta}}),
\end{equation}
where $I(\bm{\upeta};\hat{\bm{\upeta}})$ and $d(\bm{\upeta},\hat{\bm{\upeta}})$ represent the mutual information (MI) and distance metric between $\bm{\upeta}$ and $\hat{\bm{\upeta}}$, respectively. $R(D)$ characterizes the minimal information rate required to achieve a preset distortion in lossy transmission. Based on the source-channel separation theorem with distortion \cite[Th. 10.4.1]{thomas2006elements}, the recovery distortion $D_c$ at C-side is achievable if and only if $R(D_c) \le C$, where $C$ represents the capacity of communication channel. Consequently, the general optimization problem for CAS can be formulated as 
\begin{equation} \label{O1}
\begin{aligned}
\mathop { \text{min} }\limits_{\textbf{X}_s, \textbf{X}_c, D_c} \kern 5pt & D_{sc}= D_s(\textbf{X}_s)+D_c   \\
\text{subject to} \kern 5pt & R(D_c) \le C(\textbf{X}_c), \\
&\text{Tr}(\textbf{X}_s\textbf{X}_s^H)+\text{Tr}(\textbf{X}_c\textbf{X}_c^H) \le TP_T,
\end{aligned}
\end{equation}
where $P_T$ is the total power budget at BS. The first constraint is referred as to the separation theorem constraint and the second as power constraint. Although the CAS formulation presents a quite concise form, it is difficult to characterize the distortion $D_{sc}$ with the explicit expressions for arbitrary estimation tasks and prior distributions $p_{\bm{\upeta}}(\bm{\eta})$ in general. In what follows, we will focus on the specified TRM estimation as similarly considered in \cite{8579200}, to reveal the insights into CAS and leave the more general cases for future research.  

\subsection{Target Response Matrix Estimation}
In this subsection, we will derive the explicit expressions for sensing distortion $D_s$, R-D function $R(D)$ and communication channel capacity $C$ in the case of CAS-based TRM estimation task, where the parameter to be estimated is the target channel matrix itself, i.e., $\bm{\upeta}=\text{vec}(\textbf{H}_s^T)$. The vector form of sensing model (\ref{SensingM}) can be rewritten as     
\begin{equation}\label{VecModel}
\textbf{y}_s=\tilde{\textbf{X}}_s\textbf{h}_s+\textbf{z}_s,
\end{equation}  
where $\textbf{y}_s=\text{vec}(\textbf{Y}_s^T)$, $\tilde{\textbf{X}}_s=\textbf{I}_{M_s} \otimes \textbf{X}_s^T$, $\textbf{h}_s=\text{vec}(\textbf{H}_s^T)$, $\textbf{z}_s = \text{vec}(\textbf{Z}_s^T)$. Here, we make the following assumptions on the parameter's prior distribution and the random waveform

\textbf{A1}. The vector $\textbf{h}_s$ is a Gaussian random vector with the distribution $\textbf{h}_s \sim \mathcal{CN}(0,\sigma_{\eta}^2\textbf{I}_{NM_s})$.

\textbf{A2}. The Gaussian signal is used to transmit information at the C-side in order to achieve the capacity, namely, $\textbf{X}_c \sim \mathcal{CN}(0,\textbf{R}_{c_x})$, where $\textbf{R}_{c_x}$ is the covariance matrix.   

\textit{(1) Sensing Distortion:} Under the assumption \textbf{A1}, the sensing model (\ref{VecModel}) is exactly a Gaussian linear model (GLM), where the observations $\textbf{y}_s$ follows complex Gaussian distribution with zero mean and the covariance matrix
\begin{equation}
\textbf{R}_y=\sigma_{\eta}^2\tilde{\textbf{X}}_s\tilde{\textbf{X}}_s^H+\sigma_s^2\textbf{I}_{TM_s}.
\end{equation}
Thus, by adopting the MMSE estimator at S-side, the estimated  $\hat{\bm{\upeta}}_s$ can be obtained at the BS as \cite{BookEstimationTheory} 
\begin{equation}\label{ese}
\hat{\bm{\upeta}}_s = \sigma_{\eta}^2\tilde{\textbf{X}}_s^H\textbf{R}_y^{-1}\textbf{y}_s.
\end{equation}
In this case, the sensing distortion is the MMSE, expressed by
\begin{equation} \label{MMSE}
D_s(\textbf{X}_s)=\text{Tr}\left[(\frac{1}{\sigma_{\eta}^2}\textbf{I}_{NM_s}+\frac{1}{\sigma_s^2}\tilde{\textbf{X}}_s^H\tilde{\textbf{X}}_s)^{-1}\right]. 
\end{equation}

\textit{(2) R-D Function:} From (\ref{ese}), we note that the estimated parameter also follows complex Gaussian distribution with zero mean and the covariance matrix  
\begin{equation}
\textbf{R}_{\upeta_s}=\sigma_{\eta}^4\tilde{\textbf{X}}_s^H\textbf{R}_y^{-1}\tilde{\textbf{X}}_s.
\end{equation}
At the C-side, the BS transmits this Gaussian source $\hat{\bm{\upeta}}_s$ to the user and adopts the MSE as the distortion metric. This model is also referred to as the \textit{quadratic Gaussian problem}\cite{8636539}. In such case, the R-D function (\ref{ER}) has the following expression according to the result in \cite[Th. 10.3.3]{thomas2006elements}  \footnote{The coefficient $\frac{1}{2}$ is vanished since the complex variable is considered.}       
\begin{equation}\label{VRD}
R(D_c) = \sum_{i=1}^{NM_s} \log \frac{\lambda_{\upeta_i}}{D_i}, \kern 5pt 
\end{equation}
where $\lambda_{\upeta_i}$ represents the $i$-th eigenvalue of the matrix $\textbf{R}_{\upeta_s}$, the distortion $D_i$ satisfies the  inverse water-filling constraint as follows \cite{thomas2006elements}  
\begin{equation}\label{IWF}
\sum_{i=1}^{NM_s} D_i=D_c, \kern 5pt
D_i = \left\{
\begin{aligned}
	\xi, \kern 2pt \text{if} \kern 2pt \xi < \lambda_{\upeta_i}, \\
	\lambda_{\upeta_i}, \kern 2pt \text{if} \kern 2pt \xi \ge \lambda_{\upeta_i},
\end{aligned} \right.
\end{equation}
with $\xi$ being the inverse water-filling factor. 

\textit{(3) Channel Capacity:} Under the assumption \textbf{A2}, the communication channel capacity is given by
\begin{equation} \label{CD}
C=\text{log}\left|\frac{1}{\sigma_c^2}\textbf{H}_c\textbf{R}_{c_x}\textbf{H}_c^H+\textbf{I}_N\right|.
\end{equation}

\section{The Proposed Waveform Design}
Two typical signaling schemes, i.e., the separated S$\&$C and dual-functional waveform designs, have attracted extensive attention in the ISAC systems. Motivated by this, we propose the optimal waveform design approaches for the above two schemes in CAS system and reveal the corresponding performance trade-off between S- and C- sides in this section. 

For discussion convenience, we define the following sample covariance matrices of the transmitting S$\&$C waveform\footnote{The approximation between the statistical and sample covariance is accurate when the number of samples $T$ is large.} 
\begin{equation}
\tilde{\textbf{R}}_{s_x} = \frac{1}{T}\textbf{X}_s\textbf{X}_s^H, \kern 5pt \textbf{R}_{c_x} \approx \tilde{\textbf{R}}_{c_x} = \frac{1}{T}\textbf{X}_c\textbf{X}_c^H . 
\end{equation} 
The associated eigenvalue decomposition can be written by 
\begin{equation}\label{USU}
\tilde{\textbf{R}}_{s_x} = \textbf{U}_s \bm{\Lambda}_s \textbf{U}_s^H, \kern 5pt \tilde{\textbf{R}}_{c_x} = \textbf{U}_c \bm{\Lambda}_c \textbf{U}_c^H,  
\end{equation} 
where the matrices $\bm{\Lambda}_s$ and $\bm{\Lambda}_c$ are diagonal with $\lambda_{s_i}$ and $\lambda_{c_i}$ as their $i$-th diagonal element, respectively. The unitary matrices $\textbf{U}_s$ and $\textbf{U}_c$ are the corresponding eigenvector matrices.     

By utilizing formula (\ref{USU}) and the property of Kronecker operation, the sensing distortion (\ref{MMSE}) can be simplified as
\begin{equation} \label{MMSE11}
\begin{aligned}
D_s(\textbf{X}_s)&=\text{Tr}\left[\textbf{I}_{M_s} \otimes \left( \frac{1}{\sigma_{\eta}^2}\textbf{I}_{N}+\frac{1}{\sigma_s^2}\textbf{X}_s^* \textbf{X}_s^T\right)^{-1}\right]\\
&= M_s \sum_{i=1}^{N} \frac{\sigma_s^2 \sigma_{\eta}^2}{\sigma_s^2+T\sigma_{\eta}^2 \lambda_{s_i}} \mathop =  \limits^{\Delta} M_s \sum_{i=1}^{N} f_s(\lambda_{s_i}), 
\end{aligned}
\end{equation}
where $f_s(\lambda_{s_i})=\frac{\sigma_s^2 \sigma_{\eta}^2}{\sigma_s^2+T\sigma_{\eta}^2 \lambda_{s_i}}$. Similarly, the covariance matrix of the to-be-transmitted information source $\hat{\bm{\upeta}}_s$ can be rewritten as  
\begin{equation}
\begin{aligned}
\textbf{R}_{\upeta_s} &= \textbf{I}_{M_s} \otimes \left(\sigma_{\eta}^2 \textbf{X}_s^*(\textbf{X}_s^T\textbf{X}_s^*+\frac{\sigma_s^2}{\sigma_{\eta}^2}\textbf{I}_T)^{-1}\textbf{X}_s^T \right)\\
&=\textbf{I}_{M_s} \otimes \textbf{U}_s^* \bm{\Lambda}_{\upeta_s} \textbf{U}_s^T,
\end{aligned}
\end{equation}
where $\bm{\Lambda}_{\upeta_s} = \text{diag}([\lambda_{\upeta_1},\cdots,\lambda_{\upeta_N}])$ with the expression of
\begin{equation}\label{srr}
\lambda_{\upeta_i} = \frac{T\sigma_{\eta}^4\lambda_{s_i}}{\sigma_s^2+T\sigma_{\eta}^2\lambda_{s_i}} = \sigma_{\eta}^2-f_s(\lambda_{s_i}) \mathop =  \limits^{\Delta} g(\lambda_{s_i}).
\end{equation}

As for the channel capacity (\ref{CD}), the well-known optimal waveform is perfectly aligning the column space of $\textbf{X}_c$ to the eigenspace of $\textbf{H}_c^H\textbf{H}_c$, such that the MIMO channel can be decomposed into several parallel sub-channels. Specifically, we denote $\textbf{H}_c^H\textbf{H}_c=\textbf{V}_c\bm{\Lambda}_{h_c} \textbf{V}_c^H$ as the corresponding eigenvalue decomposition with $\bm{\Lambda}_{h_c}=\text{diag}([\lambda_{h_1},\cdots,\lambda_{h_N}])$. By setting $\textbf{U}_c=\textbf{V}_c$, the channel capacity (\ref{CD}) can be recast by
\begin{equation}\label{sCD}
C=\sum_{i=1}^{N}\log(\frac{T\lambda_{h_i}}{\sigma_c^2}\lambda_{c_i}+1) \mathop =  \limits^{\Delta} \sum_{i=1}^{N}\log(\alpha_i\lambda_{c_i}+1),
\end{equation}
where $\alpha_i=T\lambda_{h_i}/\sigma_c^2$.  
By substituting (\ref{VRD}), (\ref{MMSE11}), (\ref{srr}), and (\ref{sCD}) into the original optimization problem (\ref{O1}), the waveform design can be transformed into the follow form 
\begin{equation} \label{PR}
\begin{aligned}
\mathop { \text{min} }\limits_{\lambda_{s_i}, \lambda_{c_i}, D_c}  & \kern 5pt  M_s \sum_{i=1}^{N} f_s(\lambda_{s_i})+D_c   \\
\text{subject to} \kern 5pt & (\ref{IWF}), M_s \sum_{i=1}^{N} \log \frac{g(\lambda_{s_i})}{D_i} \le \sum_{i=1}^{N}\log(\alpha_i\lambda_{c_i}+1),  \\
& \sum_{i=1}^{N}\lambda_{s_i} + \sum_{i=1}^{N}\lambda_{c_i} \le P_T, \kern 5pt \lambda_{s_i}, \lambda_{c_i} \ge 0.
\end{aligned}
\end{equation}
Problem (\ref{PR}) is exactly to find a power allocation scheme to minimize the total distortion at the user under the separation theorem, inverse water-filling and power constraints. In what follows, we discuss separate and dual-functional waveform designs under the framework of (\ref{PR}).

\subsection{The Separated S$\&$C Waveform Design}
In this scenario, the BS transmits individual waveform for sensing and communication tasks, which can achieve optimal S$\&$C performance at S- and C- sides, respectively. Therefore, problem (\ref{PR}) can be divided into two sub-problems that are coupled by the total transmit power budget at the BS. Let us denote $P_s$ and $P_c$ as the power allocated to S- and C- sides satisfying $P_s+P_c=P_T$. Thus, for the given power $P_s$ at S-side, the MMSE minimization problem can be written by
\begin{equation} \label{Sopt}
\mathop { \text{min} }\limits_{\lambda_{s_1},\cdots,\lambda_{s_N}}   M_s \sum_{i=1}^{N} f_s(\lambda_{s_i}) \kern 2pt \text{subject to} \kern 2pt \sum_{i=1}^{N}\lambda_{s_i} \le P_s, \lambda_{s_i} \ge 0.
\end{equation}   
Problem (\ref{Sopt}) has the classical water-filling solution as \cite{8995606} 
\begin{equation} \label{SSopt}
\lambda_{s_1} = \cdots= \lambda_{s_N} = \frac{P_s}{N} = \xi_s.
\end{equation}   
At the C-side, since the $D(R)$ is the monotonically decreasing function with respect to $R$, minimizing the distortion $D_c$ is equivalent to maximizing the channel capacity with the following form
\begin{equation} \label{Copt}
\mathop { \text{max} }\limits_{\lambda_{c_1},\cdots,\lambda_{c_N}}   \sum_{i=1}^{N}\log(\alpha_i\lambda_{c_i}+1) \kern 2pt \text{subject to} \kern 2pt \sum_{i=1}^{N}\lambda_{c_i} \le P_c, \lambda_{c_i} \ge 0.
\end{equation} 
Again, the closed-form water-filling solution and the resulting channel capacity are given by \cite{8995606}
\begin{equation} \label{CCopt}
\lambda_{c_i} = (\xi_c-\frac{1}{\alpha_i})^+, \kern 2pt C^{\star}=\sum_{i=1}^{N}\log((\alpha_i\xi_c-1)^++1),
\end{equation}  
where $\xi_c$ is the water-filling factor such that $\sum_{i=1}^{N}\lambda_{c_i}=P_c$. Finally, we can calculate $\lambda_{\upeta_i}$ by substituting (\ref{SSopt}) into (\ref{srr}), and obtain the distortion $D_c$ for a given $C^{\star}$ through the reverse water-filling procedure (\ref{IWF}), thereby the total distortion $D_{sc}$.

Based on the above analysis, one can readily reconstruct the optimal S$\&$C waveform when the water-filling factors $\xi_s$ and $\xi_c$ are determined. Meanwhile, these two factors are coupled in the power constraint. Accordingly, problem (\ref{PR}) can be transformed into the following form  
\begin{equation} \label{SPR}
\mathop { \text{min} } \limits_{\xi_s, \xi_c}  D_{sc} \kern 5pt \text{subject to} \kern 5pt N\xi_s+\sum_{i=1}^{N}(\xi_c-\frac{1}{\alpha_i})^+ \le P_T,
\end{equation}
where $(\cdot)^+ = \text{max}\{\cdot,0\}$. Problem (\ref{SPR}) is nothing but to find how much power allocated to S- and C- sides such that the minimum distortion can be achieved at the user. Therefore, the optimal solution of (\ref{SPR}) can be found through the grid search over the power interval $[0,P_T]$, whose detailed flow is summarized in Algorithm \ref{alg1}. 

\begin{algorithm}[!t] 
	\caption{Separated Waveform Design Algorithm}   %% titile of the algorithm
	\label{alg1}
	\begin{algorithmic}		
		\STATE \textbf{Initialize:} Grid number $L$, search interval $\mathcal{U}=[0,P_T]$. \\
		\STATE \textbf{Repeat:}\\ 
		(1) Divide evenly $\mathcal{U}$ into $L$ sub-intervals with grids set $\mathcal{I}$;\\
		\textbf{For} $l=1:L-1$ \\
		(2) $P_s^{(l)} \gets \mathcal{I}(l)$, $P_c^{(l)} \gets P_T-P_s^{(l)}$; \\
		(3) Calculate $D_s^{(l)}$ according to (\ref{SSopt}) and (\ref{MMSE11}) with $P_s^{(l)}$ ; \\
		(4) Calculate $C^{(l)}$ according to water-filling (\ref{CCopt}) with $P_c^{(l)}$; \\
		(5) Calculate $D_c^{(l)}$ according to reverse water-filling (\ref{IWF}) with $C^{(l)}$, and $D_{sc}^{(l)} \gets D_s^{(l)}+D_c^{(l)}$;\\
		\textbf{end} \\
		(6) Find $P_s^{(l^*)} = \mathop { \text{min} } \limits_{P_s^{(l)} \in \mathcal{I}}  D_{sc}^{(l)} $, and $\mathcal{U} \gets [P_s^{(l^*-1)},P_s^{(l^*+1)}]$;
		\STATE \textbf{Until:} the algorithm convergence.	
	\end{algorithmic}
\end{algorithm}

\textit{Remark 1 (Power Allocation Trade-off)}: The separated waveform design demonstrates a clear power allocation trade-off between S- and C- sides. Evidently, excessively allocating resources to either side will cause the sensing QoS significant degradation at the user. Although only the power allocation is considered in this paper, the system resources including bandwidth, multi-beams, dwell time, etc., may also affect the CAS performance, which is left for our future research.       

\subsection{The Dual-functional Waveform Design}
In this subsection, we focus on the dual-functional waveform applied for both S- and C-sides, i.e., $\textbf{X}_s=\textbf{X}_c=\textbf{X}$. This corresponds to that the BS transmits the parameter data of the last epoch at C-side and uses the same waveform for sensing at the current epoch. We should highlight the following essential difference compared to the above separated waveform design.
   
%$\bullet$ \textit{Deterministic-random trade-off}: For communication, one prefers the entropy of $\textbf{X}$ as large as possible to convey more information. In contrast, the deterministic signals are required for achieving a stable sensing performance. This leads to the \textit{deterministic-random} trade-off in dual-functional waveform design \cite{xiong2022flowing}. Consequently, the assumption $\textbf{A2}$ may cause the performance degradation for sensing. 

\textit{Remark 2 (Water-filling Trade-off)}: We can observe that the sensing-optimal waveform prefers to uniformly allocate the power in (\ref{SSopt}), whereas the communication-optimal waveform needs to allocate power in terms of the channel $\textbf{H}_c$ through water-filling method in (\ref{CCopt}). One cannot find a dual-functional waveform that simultaneously optimizes both S- and C- sides with high probability, unless the sensing and communication channels are sufficiently similar. This leads to a water-filling trade-off.   

By transforming the reverse water-filling (\ref{IWF}) into closed-form, the original problem for dual-functional waveform design can be recast by  
\begin{equation} \label{DFCAS} 
\begin{aligned}
\mathop { \text{min} }\limits_{\tilde{\lambda}_i, \xi} \kern 5pt & \sum_{i=1}^{N} M_s f_s(\tilde{\lambda}_i) + g(\tilde{\lambda}_i) - \left(g(\tilde{\lambda}_i)-\xi\right)^+ \\
\text{s.t.} \kern 5pt
&M_s\sum_{i=1}^{N}\log\frac{g(\tilde{\lambda}_i)}{g(\tilde{\lambda}_i) - \left(g(\tilde{\lambda}_i)-\xi\right)^+} \le \sum_{i=1}^{N} \log (\alpha_i\tilde{\lambda}_i+1),\\
&\sum_{i=1}^{N} \tilde{\lambda}_i \le P_T, \kern 2pt, \tilde{\lambda}_i \ge 0,  
\end{aligned}
\end{equation}
where the variables are transformed into the shared use power allocation $\tilde{\lambda}_i$ and inverse water-filling factor $\xi$. This non-convex problem is difficult to cope with in general due to the inverse water-filling procedure and the fact that the variables are coupled in the separation theorem constraint. To tackle this problem, we present a heuristic sub-optimal search algorithm. The core idea is setting the sensing (or communication) optimal solution as the initial point, gradually re-allocating the power to improve the communication (or sensing) performance, until the objective function is minimized.     

Initially, we set the sensing-optimal scheme $\bm{\lambda}^\text{s-opt}$ or communication-optimal scheme $\bm{\lambda}^\text{c-opt}$ as the initial point. The $\bm{\lambda}^\text{s-opt}$ and $\bm{\lambda}^\text{c-opt}$ are the solutions of problem (\ref{Sopt}) and (\ref{Copt}) where the power budget is replaced by the total power $P_T$, respectively. Let us take $\bm{\lambda}^\text{(0)}=\bm{\lambda}^\text{s-opt}= [\tilde{\lambda}^{(0)}_i,\cdots,\tilde{\lambda}^{(0)}_N]^T$ as an example. The distortion $D_{sc}^{(0)}$ can be readily obtained by following the similar procedure in the separated scheme. Then, the search direction is chosen to improve the communication performance, i.e.,
\begin{equation}
\nabla C(\bm{\lambda})=\left[\frac{\alpha_1}{\alpha_1\tilde{\lambda}_1+1},\cdots,\frac{\alpha_N}{\alpha_N\tilde{\lambda}_N+1}\right]^T. 
\end{equation}      
Thus, the solution is updated by
\begin{equation}\label{search1}
\bm{\lambda}^\text{(1)} = \bm{\lambda}^\text{(0)} + \beta\nabla C(\bm{\lambda}^\text{(0)}), \kern 5pt \bm{\lambda}^\text{(1)} = \frac{\bm{\lambda}^\text{(1)}}{\|\bm{\lambda}^\text{(1)}\|_1}P_T. 
\end{equation}    
where the second formula in (\ref{search1}) is to meet the power constraint. For a given solution $\bm{\lambda}^\text{(1)}$, we can update the distortion $D_{sc}^{(1)}$ in the new iteration. Furthermore, the step size $\beta$ is chosen by the classical backtracking line search, ensuring the value of objective $D_{sc}$ non-increasing at each iteration. Therefore, the iterative procedure is stopped until $D_{sc}$ remains constant. In summary, the proposed dual-functional waveform design algorithm can be found in Algorithm \ref{alg2}.  

\begin{algorithm}[!t] 
\caption{Separated Waveform Design Algorithm}   %% titile of the algorithm
\label{alg2}
\begin{algorithmic}		
\STATE \textbf{Initialize:} $\bm{\lambda}^\text{(0)} = \bm{\lambda}^\text{s-opt}, D_{sc}(\bm{\lambda}^\text{(0)})$, $\epsilon$. \\
\STATE \textbf{Repeat:}\\ 
(1) Calculate $\bm{\lambda}^\text{(1)}$ by (\ref{search1});   \\
(2) Calculate $D_{sc}(\bm{\lambda}^\text{(1)})$ similar as Algorithm \ref{alg1}; \\
(3) Update $\bm{\lambda}^\text{(0)} \gets \bm{\lambda}^\text{(1)}$; \\
\STATE \textbf{Until:} $|D_{sc}(\bm{\lambda}^\text{(1)})-D_{sc}(\bm{\lambda}^\text{(0)})| \le \epsilon$.	
\end{algorithmic}
\end{algorithm} 

\section{Simulation Results}
In this section, we evaluate the effectiveness of the waveform design schemes for the CAS systems. The BS is equipped with $N=10$ transmitting antennas, and the receiving antennas at S- and C- sides are set as $M_s=M_c=5$. We set a relatively large number of symbols $T=100$ to reduce the approximate error between statistical and sample covariance matrices. At C-side, the Rayleigh fading model for the communication channel is adopted, where each entry of $\textbf{H}_c$ obeys the standard complex Gaussian distribution. In contrast, the sensing channel covariance factor is set as $\sigma_{\eta}^2=0.1$. Without loss of generality, we use the normalized transmit power $P_T = 1W$ and define the signal to noise ratio (SNR) by $\text{SNR}=10\log TP_T/\sigma^2_{s(c)}$ dB. In the following simulations, we fix the SNR for sensing channel by setting $\text{SNR}_s=20$dB, and evaluate impacts of the SNR for communication channel on the sensing QoS (i.e., distortion $D_{sc}$) at the user.   

\begin{figure}[!t]
	\centering
	\includegraphics[width=3in]{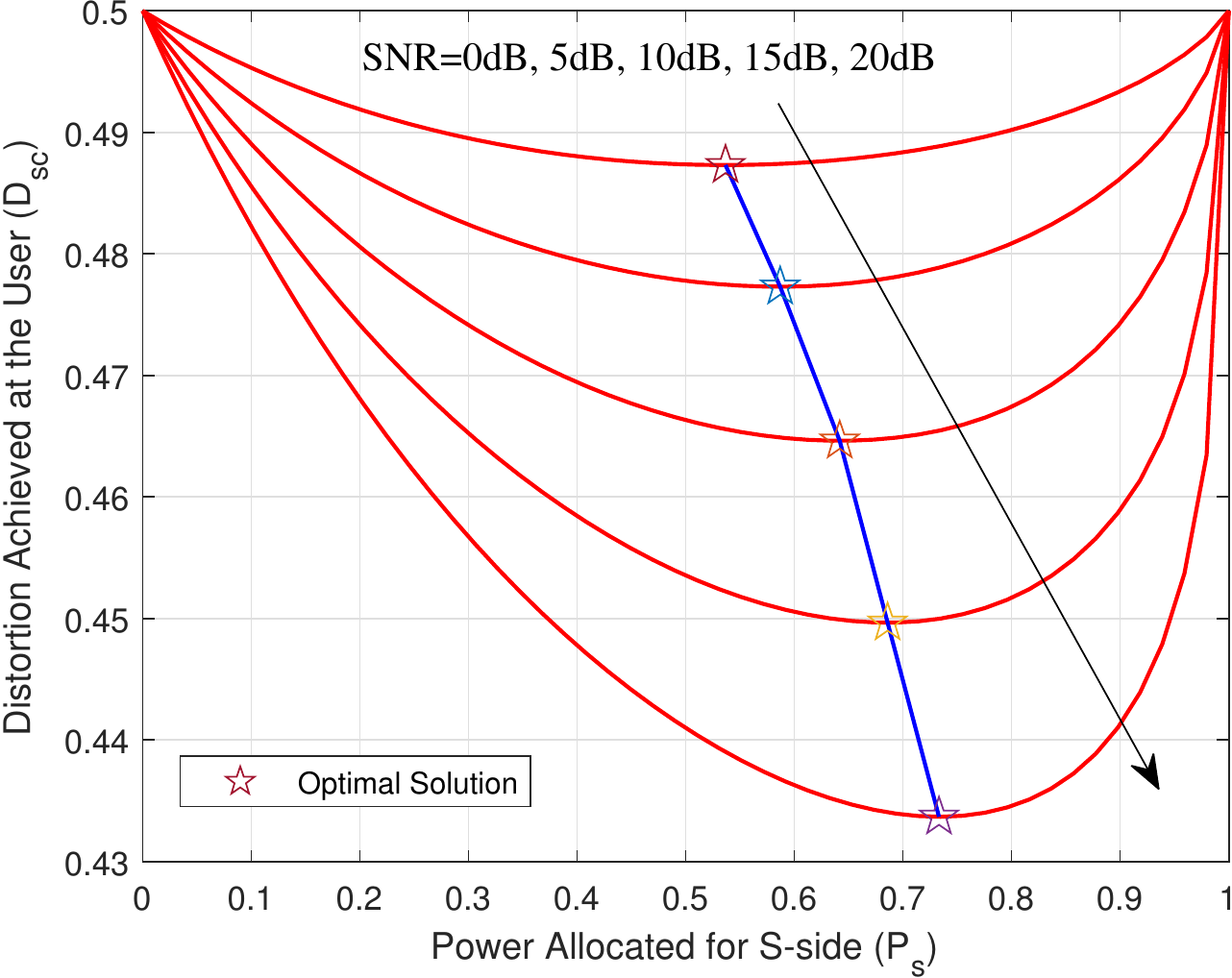}
	\caption{The distortion versus power allocation for separated waveform design.}
	\label{SWD}
\end{figure}

First, we provide the distortion achieved at the user versus the power allocation for separated waveform design, while varying the SNR between 0dB and 20dB. As shown in Fig. \ref{SWD}, the proposed Algorithm \ref{alg1} effectively obtains the optimal solution within the power allocation curves. Moreover, the figure depicts that, as the SNR levels at the C-side increase, more power is allocated for the S-side, resulting in a performance trade-off between the S and C-sides. Notably, the quality of communication channel plays an crucial role in the CAS systems. This implies that the sensing QoS can only be enhanced with a sufficiently good communication channel.

\begin{figure}[!t]
	\centering
	\includegraphics[width=3.1in]{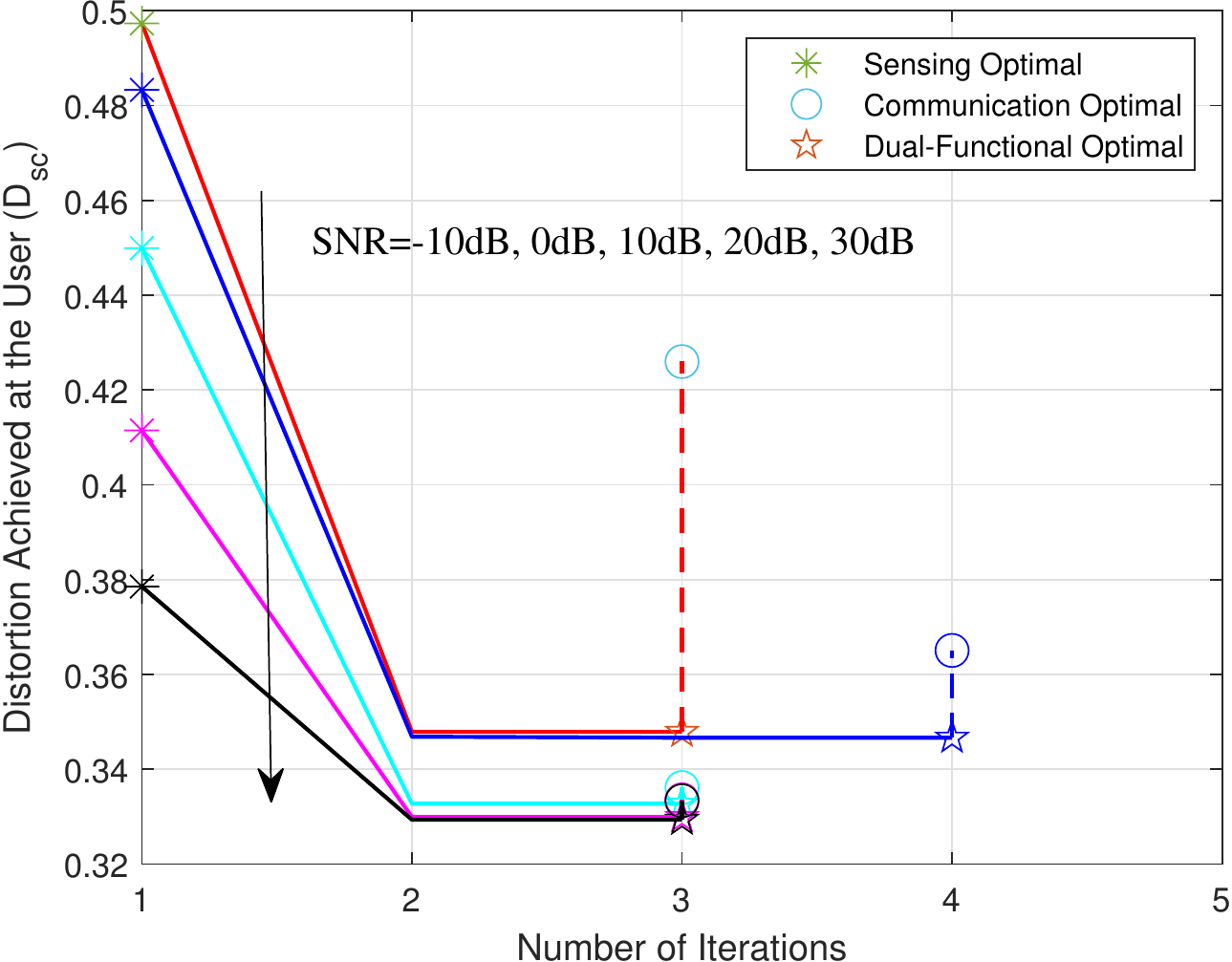}
	\caption{The distortion versus number of iterations for dual-functional waveform design.}
	\label{DFD}
\end{figure}
\begin{figure}[!t]
	\centering
	\includegraphics[width=3in]{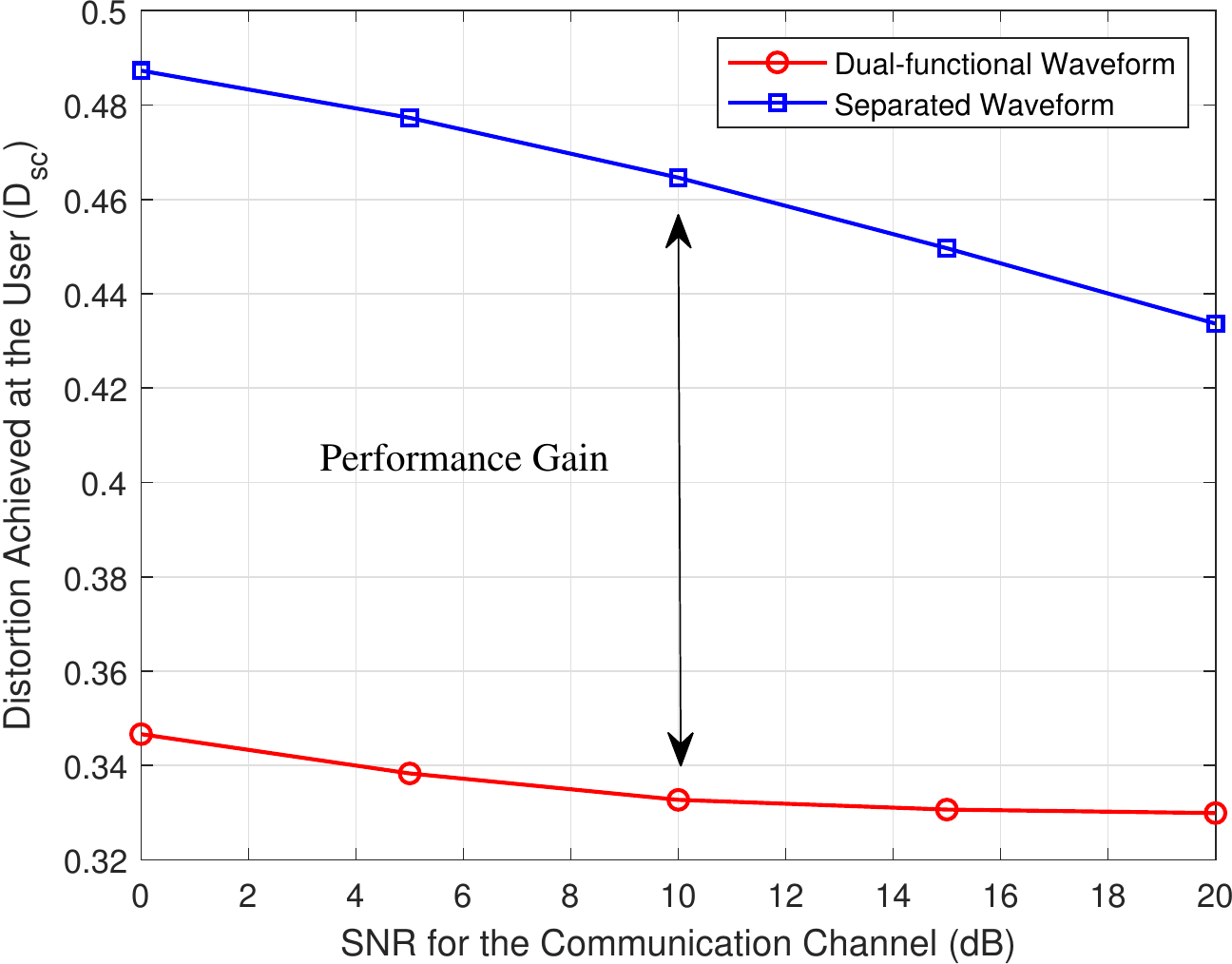}
	\caption{The performance comparison between separated and dual-function waveform designs.}
	\label{CompareSD}
\end{figure}

Next, we evaluate the sensing QoS achieved by the dual-functional waveform design. Fig. \ref{DFD} illustrates that iterative Algorithm \ref{alg2} ensures convergence to a sub-optimal solution within a few iterations. The sensing- and communication- optimal schemes refer to minimizing the MMSE at S-side and maximizing the capacity at C-side, respectively. At the regime of low SNRs (SNR$<$0dB), we can identify compromise power allocation schemes that outperform both sensing- and communication- optimal schemes. Subsequently, as SNRs increase, the solution tends to attain the optimal communication performance. Conversely, once the communication channel is adequately good, the waveform prioritizes improving the sensing performance as much as possible, resulting in a sensing-optimal scheme. Similar to the separated waveform design, a better communication channel gains a lower sensing distortion at the user.      

Finally, we compare the performance of the proposed two schemes under the same system settings. While the other resource allocation aspects aside from power and the mutual interference caused by the separated waveform are omitted in this paper, Fig. \ref{CompareSD} indicates that the dual-functional design attains approximately a 25\% performance gain compared to the separated design thanks to the shared use of the transmit power budget at the BS. 

\section{Conclusion}\label{Conclusion}
In this article, we presented a novel framework for waveform design in communication-assisted sensing (CAS) toward the 6G perceptive network. We proposed two transmission schemes, separated sensing and communication (S$\&$C) and dual-functional waveform designs, which aim to minimize the sensing distortion at the user while satisfying the constraints of separation theorem and power budget. Based on that, We discussed the trade-off involved in power allocation for the separated S$\&$C waveform design, and the water-filling trade-off for the dual-functional waveform design. Our numerical simulations confirmed the effectiveness of the proposed algorithms, and showed a 25\% performance gain attained by the dual-function scheme compared to the separated scheme.

\bibliographystyle{IEEEtran}
\bibliography{IEEEabrv,Com_Assisted_Sensing}

\end{document}